\documentclass[journal]{IEEEtran}
\usepackage[latin9]{inputenc}
\usepackage{array}
\usepackage{verbatim}
\usepackage{amsmath}
\usepackage{amssymb}
\usepackage[square, numbers]{natbib}
\usepackage{hyperref}
\hypersetup{
 pdfauthor={Reham�Badawy^{1*},� Yordan�Raykov^{1*}� and�Max�A. Little^{1,2}�*These authors contributed equally to this work.1. Aston University, UK; 2. Media Lab, Massachusetts Institute of Technology, US �}}
\usepackage{breakurl}

\makeatletter

\providecommand{\tabularnewline}{\\}

\ifCLASSINFOpdf
\else
\fi
\usepackage{url}

\usepackage{multirow}
\usepackage{subfigure} 
\usepackage{graphicx} 
\usepackage{colortbl}
\definecolor{grey}{RGB}{210, 210, 210}
\DeclareMathOperator*{\argmax}{arg\,max}

\makeatother

\begin{document}

\title{A unified algorithm framework for quality control \\
 of sensor data for behavioural clinimetric testing\\
{\large{}Reham~Badawy}\thanks{{*}These authors contributed equally to this work.}{\large{}$^{1*}$,
Yordan Raykov$^{1*}$, Max A. Little$^{1,2}$}\thanks{1. Aston University, UK; 2. Media Lab, Massachusetts Institute of
Technology, US}}
\maketitle
\begin{abstract}
The use of smartphone and wearable sensing technology for objective,
non-invasive and remote clinimetric testing of symptoms has considerable
potential. However, the clinimetric accuracy achievable with such
technology is highly reliant on separating the useful from irrelevant
or confounded sensor data. Monitoring patient symptoms using digital
sensors outside of controlled, clinical lab settings creates a variety
of practical challenges, such as unavoidable and unexpected user behaviours.
These behaviours often violate the assumptions of clinimetric testing
protocols, where these protocols are designed to probe for specific
symptoms. Such violations are frequent outside the lab, and can affect
the accuracy of the subsequent data analysis and scientific conclusions.
At the same time, curating sensor data by hand after the collection
process is inherently subjective, laborious and error-prone. To address
these problems, we report on a unified algorithmic framework for automated
sensor data quality control, which can identify those parts of the
sensor data which are sufficiently reliable for further analysis.
Algorithms which are special cases of this framework for different
sensor data types (e.g. accelerometer, digital audio) detect the extent
to which the sensor data adheres to the assumptions of the test protocol
for a variety of clinimetric tests. The approach is general enough
to be applied to a large set of clinimetric tests and we demonstrate
its performance on walking, balance and voice smartphone-based tests,
designed to monitor the symptoms of Parkinson's disease. 

\end{abstract}

\section{Introduction}

\IEEEPARstart{I}{n} recent years sensors embedded in smartphones
and wearable devices have become ubiquitous, and have evolved to the
point where they can be used in areas such as healthcare \citep{dai2010perfalld,sha2008spa,oliver2007healthgear},
environmental monitoring \citep{maisonneuve2009noisetube,mun2009peir}
and transport \citep{thiagarajan2009vtrack}. In healthcare, for example,
smartphone sensors have been successful at detecting the symptoms
of neurological disorders such as Parkinson\textquoteright s disease
(PD). PD is a brain disease that significantly affects voluntary movement.
Symptoms of PD include slowness of movement (bradykinesia), trembling
of the hands and legs (tremor), absence of movement and loss of balance
(postural instability). Through a smartphone application, on-board
sensors in the smartphone capture the behaviour of the user while
they carry out a simple clinimetric test protocol, such as walking
in a straight line with the smartphone in their pocket \citep{arora2014high},
to detect the key symptoms of the disease. Collecting objective symptom
measurements with clinimetric testing performed on technologies such
as smartphones \citep{joundi2011rapid,kostikis2014smartphone}, or
portable and wearable sensors \citep{andrzejewski2016wearable,rigas2012assessment,Patel2009},
eliminates much of the subjective bias of clinical expert symptom
measurement, while also allowing for remote, long-term monitoring
of patient health \citep{maetzler2013quantitative}; in contrast to
the current ``snapshot'' in time obtained during a clinical visit.
Thus, remote, long-term monitoring allows for improved analysis of
a patient\textquoteright s health and outcomes. Usually, data from
such sensors is collected and analysed under a set of clinimetric
test protocol assumptions, such as the type of behaviour to be carried
out to probe for specific symptoms. Getting the assumptions of a test
protocol to hold outside controlled lab settings is a universal problem,
since uncontrollable confounding factors in the environment, such
as unavoidable and unexpected behaviours, can have an adverse impact
on the measurement process.

When the test protocol assumptions do not hold for the sensor data,
the interpretation of the analysis results becomes dubious \textendash{}
we are unlikely to be analysing the behaviour we believe we have captured
for subsequent symptom measurement. Moreover, analysing confounded
or contaminated data produces misleading, biased results which are
inherently non-reproducible and non-replicable \citep{fan2014challenges}.
In many consumer applications which use sensor technologies, such
data collection quality issues may not be that important, but they
are of critical importance in the medical sciences. Non-reproducible
results in clinimetric studies could have significant implications
for an individual\textquoteright s health. 

\begin{figure*}[t]  
\center {
\includegraphics[width=2.065\columnwidth]{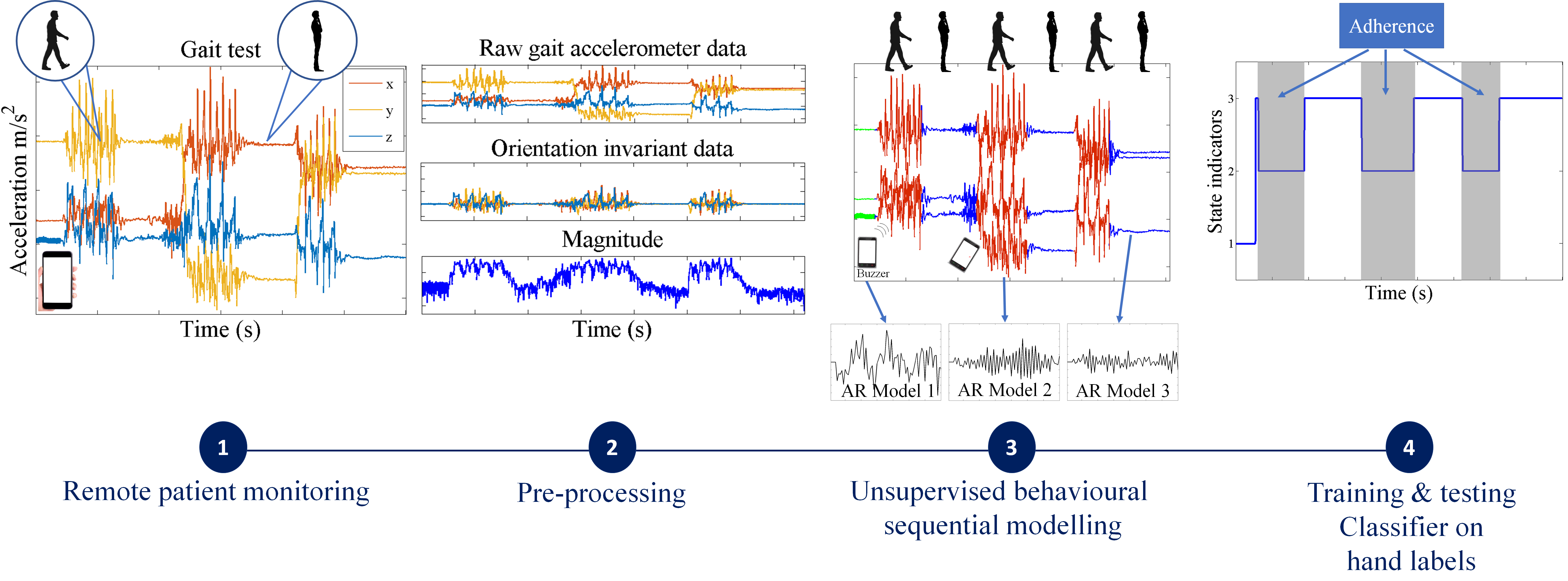}

}
\caption{Overview of the proposed algorithmic framework for data quality control of sensor data for behavioural clinimetric testing. The first stage (starting from the left) involves collecting sensor data outside the lab. The second pre-processing stage consists of removing confounding factors from the data such as the effect of orientation of the device. The third stage involves unsupervised behavioural segmentation of the data into intervals in which the user performs similar activities. In the final stage, a simple, interpretable classifier is trained to predict which behavioural intervals are associated with adherence to, and which with violations of, the test protocols.}
\label{fig.Overview-chart}
\end{figure*}

In clinimetric testing, test protocols are comprised of specific activities
(behaviours) that a user is required to perform. This means that the
quality control process can be viewed as the problem of locating different
user behaviours and assessing if those behaviours are in adherence
with the protocol assumptions. Yet it is not feasible to approach
this problem using methods used for ``activity recognition'', for
the following reasons: typically, in activity recognition in ``ubiquitous
computing'' applications, the sensor data is segmented into windows
of fixed alignment and equal duration and then a hand-crafted set
of features is extracted from each window \citep{bulling2014tutorial}.
Subsequently, these features are used to train a classification algorithm
that predicts the activity in each window. One of the problems is
that both the hand-engineering of features and the training of the
classifier in such systems depends heavily on having detailed, labeled
information about which activities actually occurred. However, outside
the lab such information is rarely available. The second problem with
current approaches to activity recognition, is that usually they rely
on modeling the feature space instead of the raw data \citep{zhang2017fault}.
This means that often they do not take into account the temporal dependence
of the sensor data between windows. These limitations are not reflected
in the reported accuracies of these system since they are trained
and tested only under controlled lab environments \citep{borve2015smartphone,stedtfeld2012gene,odeh2015optimizing,Patel2009}.

Performance of heavily ``fine-tuned'' machine learning systems for
activity recognition are misleading if the sensor data collected outside
the lab is drawn from a different distribution to that collected in
the lab and used for training the system \citep{hand2006}. This problem
is compounded when using high performance nonlinear classification
algorithms (such as \textit{convolutional neural networks}, \textit{random
forest classifiers} or \textit{support vector machines}) on a large
number of features all estimated from a training distribution of questionable
relevance in practice.

To address these issues, in this study we propose a unified algorithm
framework for automated assessment of clinimetric sensor data quality,
i.e. the extent to which the data adheres to the assumptions of the
clinimetric test protocols. Combining both parametric and nonparametric
signal processing and machine learning techniques we demonstrate the
scope, effectiveness and interpretability of this framework by applying
it to multiple sensor types and clinimetric tests for assessment of
PD. Across 100 participants and 300 clinimetric tests from 3 different
types of behavioural clinimetric protocols, the system shows average
segmentation accuracy of around 90\% when compared to a human expert
performing the same quality control task manually. We focus on data
collected from smartphone sensors deployed outside the lab, as these
are the most ubiquitous devices available for objective symptom measurement
in practice.

\section{Related work}

Smartphones and wearables are increasingly recognized as potential
tools for remote monitoring, diagnosis and symptom assessment of patients
with various conditions: wearable wrist bands have been increasingly
used to predict epileptic seizures \citep{poh2012convulsive}; smartphones
have proven accurate for monitoring of PD symptoms in clinical trials
\citep{lipsmeier2017successful}; shoe-based devices have been developed
to support rehabilitation of patients who have suffered through a
stroke \citep{edgar2010wearable}; and apps have been developed that
provide insulin dosage recommendations for type 1 diabetes patients
\citep{charpentier2011diabeo}. Many other recent healthcare applications
of consumer electronic devices can be found in wider reviews such
as \citet{ozdalga2012smartphone}, \citet{mosa2012systematic} and
\citet{kubota2016machine}. For such devices to become useful clinical
tools they have to be deployed in realistic natural environments such
as the private home or the office. However, it is difficult to ensure
that the same machine learning systems that have been developed in
a controlled lab setting can perform well in unknown environmental
settings. Training and testing systems on data collected outside of
the lab is not really feasible since we have very little labeled information
about detailed user behaviours in this situation. Such labels are
commonly generated by manual hand labeling by a trained expert, video
monitoring of users and user self-assessment. Video recordings are
subsequently manually annotated and common in continuous monitoring
systems, for example tremor detection \citep{garcia2016} or PD disease
state assessment \citep{hammerla2015pd}. Using video annotations
significantly complicates the experimental setup, makes clinimetric
studies a lot more expensive and is still quite difficult to analyze.
This is why often only certain parts of the video recording are used
to obtain labels for parts of the collected data. Alternatively, self-assessment
and self-report diaries usually deviate significantly from expert
assessment, at least for neurological disorders such as PD \citep{Reimer2004b}.
Manual expert annotation of sensor data is also not always objective
and typically can provide only broad indications of user behaviour
or health status. Unfortunately, these issues are often overlooked,
in particular when it comes to the evaluation of clinimetric testing
tools and studies often rely only on data captured in the lab \citep{hoff2001accelerometric,cole2010,giuffrida2009clinically}.

Assuming only some limited amount of labeling is available, in this
paper we propose an automated system which aims to recognize deviation
between the instructions of the clinimetric test protocol and the
user behaviour, in realistic settings outside the lab environment.
Since clinimetric test protocols are expressed in terms of assumptions
about the user's behaviour, the quality control algorithms we develop
here are related to, but quite different from, existing activity recognition
systems. Activity classification frameworks have been used in PD symptom
assessment systems \citep{zwartjes2010,salarian2007ambulatory,tzallas2014perform}.
For example, \citet{zwartjes2010} and \citet{salarian2007ambulatory}
developed an in-home monitoring system that detects specific behaviours,
and subsequently predicts the movement impairment severity of these
activities in terms of common PD symptoms. Both studies use existing
activity classification methods which rely upon training on a predefined,
specific set of motion-related features which can be used to distinguish
between a selected, fixed set of activities. However, it is not feasible
in practice to anticipate the entire behavioural repertoire of a participant
during any clinimetric test conducted outside the lab. Existing activity
recognition systems require a rich set of features to be extracted
from the input data and the choice of features depends on the activities
we wish to discriminate between \citep{bulling2014tutorial}. For
these reasons most traditional supervised activity recognition systems
are not feasible for quality control in the context discussed in this
paper, where any set of a potentially infinite range of behaviours
could be encountered. A solution to tackling the wide range of possible
behaviours in practice is to cluster the data into variable size windows
using segmentation, where the specific activity in each segment is
not specified. \citet{guo2012} proposed a somewhat more adaptive
segmentation approach which does not rely on an a-priori fixed set
of features, but uses \textit{\emph{principal component analysis}}
to select the most appropriate features for the different segmentation
tasks. The segmentation of activities is then mostly performed using
bottom-up hierarchical clustering on windows of 5 to 10 seconds duration.
\citet{hammerla2015pd} also proposed an approach less reliant on
activity-specific features which utilizes a deep belief network with
two-layer restricted Boltzmann machines. However, the deep network
itself is trained on generic features extracted from 1 minute windows
of sensor data, therefore it cannot be used to capture short-term
deviations from test protocols.

To design a robust system which can detect short-term deviations from
test protocols without relying on hard-to-obtain, detailed labels
about the user behaviour during multiple tests, we turn to an unsupervised
segmentation approach. Unsupervised time series segmentation methods
have been widely studied across many disciplines including: computer
vision and graphics \citep{zhang2017fault,zelnik2006statistical,lu2004repetitive,Zhou2013};
data mining \citep{keogh2001online}; speech recognition \citep{shannon1995speech,Rabiner1989}
and signal processing \citep{chiang2008hidden,li2007efficient}. However,
most of those techniques cannot be readily applied to low-dimensional
sensor data since their usefulness relies on a significant number
of domain-specific features. 

\section{Overview}

In this section we describe the stages in our proposed unified framework
for quality control of clinimetric test sensor data (Figure \ref{fig.Overview-chart}).
As an example application we apply the system to multiple clinimetric
tests for PD symptom monitoring. We describe the test protocols for
these clinimetric tests in Section \ref{sec:Experimental-setup}. 

In the first stage, we apply practical preprocessing steps which,
depending on the type of sensor produce a more compact representation
of the data without discarding any essential structure. For example,
in the case of accelerometer data from sensors embedded in smartphones,
we can remove the effect of orientation changes of the smartphone.
This is because device orientation is usually a confounding factor
in clinimetric testing. For high sample-rate voice data we segment
the original signal into short duration, 10ms windows and extract
features such as the energy or spectral power in each window, instead
of modeling the raw data directly. Unlike most of the existing machine
learning strategies for processing sensor data we do not rely on a
large number of features extracted from each sensor type; we apply
minimal transformations to the raw data keeping it at relatively high
sampling frequency, and directly fit a flexible probabilistic model
aiming to capture the structure of importance to the problem of quality
control.

Once the sensor data has been preprocessed accordingly, we fit in
an unsupervised manner a discrete latent variable model to each of
the sensor signals and use it to split the data into segments of varying
duration. Depending on the complexity of the data produced by different
clinimetric tests, we propose two different segmentation models which
vary in terms of flexibility and computational simplicity:
\begin{enumerate}
\item For simpler quality control problems, we develop a Gaussian mixture
model (GMM) based approach which attempts to cluster the raw signal
into two classes: data adhering or not adhering to the test protocols.
Since GMMs ignore the sequential nature of the sensor data, we pass
the estimated class indicators through a running median filter to
smooth out unrealistically frequent switching between the two classes. 
\item We also propose a more general solution which involves fitting flexible
nonparametric switching autoregressive (AR) models to each of thee
preprocessed sensor signals. The switching AR model segments the data
in an unsupervised way into a random (unknown) number of behavioural
patterns that are frequently encountered in the data. An additional
classifier is then trained to discriminate which of the resulting
variable-length segments represent adherence or violation of the test
protocol. We demonstrate that a simple multinomial naive Bayes classifier
can be trained using a strictly limited amount of labeled data annotated
by a human expert. Since the instructions in any clinimetric test
protocol are limited, whereas the number of potential behavioural
violations of the protocol are not, we assume that any previously
unseen segments that we detect have to be a new type of violation
to the specified instructions of the protocol. We have detailed the
segmentation and classification process in Section \ref{sec:Sequential-behaviour-modeling}.
\end{enumerate}

\section{Experimental setup\label{sec:Experimental-setup}}

\subsection{Data collection}

\noindent To illustrate our novel framework in practice, we use data
from the Smartphone-PD study \citep{zhan2016high}, which utilizes
an Android OS smartphone application to capture raw sensor data from
the digital sensors embedded in the device. The application prompts
the user to undertake short (less than 30 seconds) self-administrated
clinimetric tests designed to elicit the symptoms of PD. These tests
are: (1) voice test (microphone) which measures impairment in the
production of vocal sounds; (2) balance test (accelerometer) which
measures balance impairment (postural instability); and (3) walk test
(accelerometer) which measures impairment in a user's walking pattern
(Table \ref{tab:Test-protocol-and assumptions}).

\begin{table*}
\caption{\label{tab:Test-protocol-and assumptions}}
Clinimetric test protocols, assumptions and commonly encountered protocol
violations in the Smartphone-PD project data used to test the quality
control framework proposed in this study.
\par
\begin{centering}
\smallskip{}
\par\end{centering}
\centering{}%
\begin{tabular}{|>{\raggedright}p{0.9cm}|>{\raggedright}p{3.5cm}|>{\raggedright}p{3.75cm}|>{\raggedright}p{5.5cm}|}
\hline 
\textbf{Test} & \textbf{Protocol}  & \textbf{Protocol violations} & \textbf{Hand-labelling protocol}\tabularnewline
\hline 
Voice  & \raggedright{}Place the phone up to your ear as if making a normal
phone call. Take a deep breath, and say \textquotedblleft aaah\textquotedblright{}
for as long as you can, at a steady volume and pitch. & \begin{raggedright}
1. User interactions with the smartphone including taking a phone
call, texting or playing a game during test.
\par\end{raggedright}
\begin{raggedright}
2. User performs test in loud environment. 
\par\end{raggedright}
\raggedright{}3. Non-sustained vowel phonation activities including
coughing, reading out the instructions on the display, talking to
another person, during test. & (1) Vowel sound segments are marked as adherence

(2) Anything else is marked as non-adherence\tabularnewline
\hline 
Balance  & Place the phone in your pocket. When the buzzer vibrates, stand up
straight unaided. & 1. User interactions with the smartphone including taking a phone
call, texting or playing a game during test. 

2. User is jumping or falling during the test. & (1) Where there is uncertainty in the user\textquoteright s activity,
a non-adherence label is applied

(2) Buzzer is labelled as non-adherence

(3) Where an interval is confounded with the buzzer, a non-adherence
label is given to that interval

(4) Where an interval is confounded with an orientation change of
the smartphone and the data is otherwise ambiguous, a non-adherence
label is given to that interval\tabularnewline
\hline 
Walking & Stand up and place the phone in your pocket. When the buzzer vibrates,
walk forward 20 yards; then turn around and walk back. & 1. User interactions with the smartphone including taking a phone
call, texting or playing a game during test. 

2. Non-walking activities including jumping, falling or standing still.

3. User encounters obstacles during walking which interfere with normal
walking. & (1) Where there is uncertainty in the user\textquoteright s activity,
a non-adherence label is applied

(2) Buzzer is labelled as non-adherence

(3) Where an interval is confounded with the buzzer, a non-adherence
label is given to that interval

(4) Where an interval is confounded with an orientation change of
the smartphone and the data is otherwise ambiguous, a non-adherence
label is given to that interval

(5) A turn is labelled as adherence\tabularnewline
\hline 
\end{tabular}
\end{table*}

\subsection{Hand-labelling for algorithm evaluation\label{subsec:Hand-labelling-for-algorithm}}

\noindent In order to evaluate the performance of the automated quality
control algorithms developed here, some reference data is needed.
To this end, the data from some of the selected Smartphone tests were
hand-labeled according to whether it represents behaviour that adheres
to the test protocol or violates it (those labels will later appear
as $u_{1},\dots,u_{T}$ which take value $1$ for adherence and value
$2$ for violation). Note that this is an inherently subjective process
and we cannot be sure of the exact activity occurring during any period
of time. However, the hand-labelling at least provides an example
of how a human expert would classify the collected data. Thus, our
aim is not to create an algorithm which blindly reproduces the hand
labels as they are imperfect. Instead, we aim to develop an approach
that learns the major structural differences between data violating,
and data adhering to, test protocols.

We labelled data from 100 subjects (voice, balance and walking tests)
from the Smartphone-PD data, randomly selecting 50 PD patients (25
males and 25 females) and 50 healthy controls (25 males and 25 females).
Subjects are age and gender-matched (two-sample Kolmogorov-Smirnov
test%
) to rule out potential age or gender confounds (Table \ref{tab:Test-protocol-and assumptions}).

\begin{figure}[t]

\vspace{0.2cm}
\includegraphics[width=1\columnwidth]{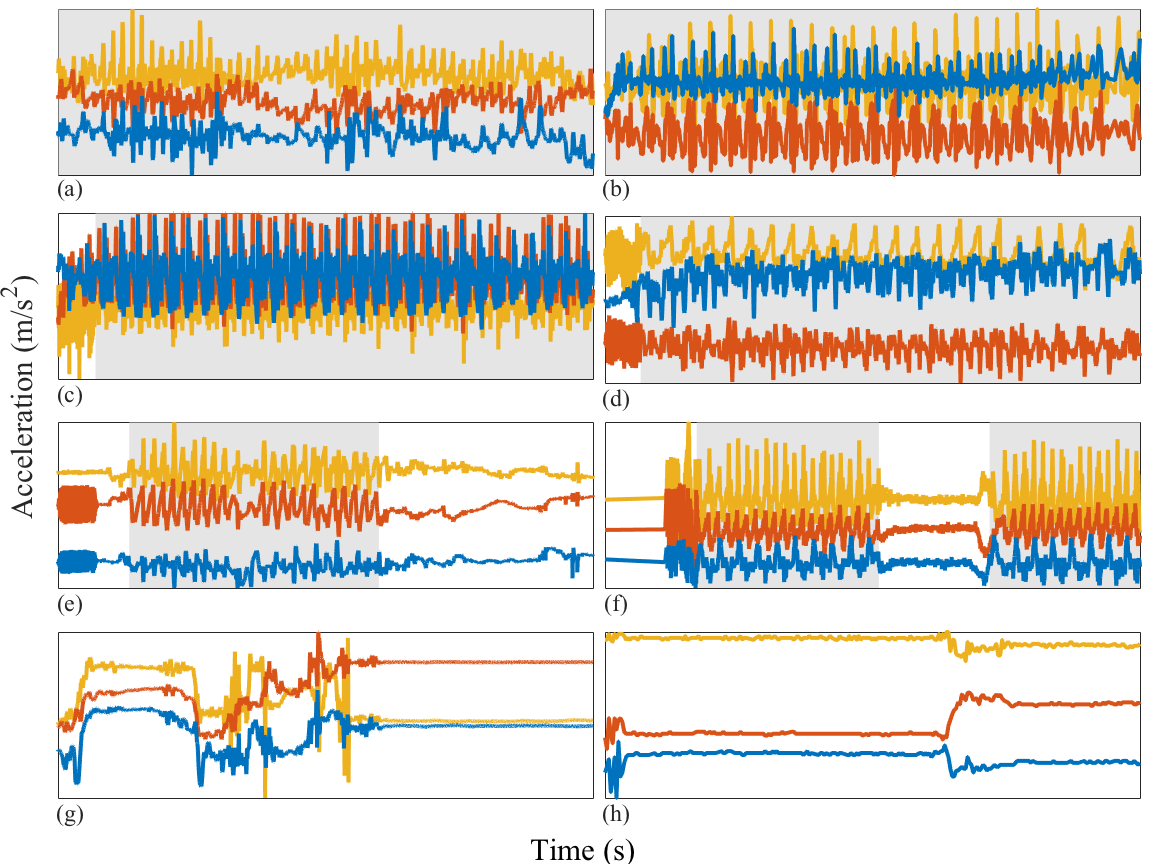}

\medskip 
\small
\caption{Illustrative examples of walking clinimetric test sensor data recorded using a smartphone accelerometer from healthy individuals and those with Parkinson's disease. 
Left column: walking tests from PD patients, right column: healthy individuals. Horizontal axis: time in seconds (approximately 30s), vertical axis (orange, red and blue) represents acceleration (m/s). Grey shaded areas: data segments in which the user is hand-labelled as adhering to the test protocol. (a) PD patient walking throughout test, (b) healthy individual walking throughout test, (c) smartphone buzzer in first few seconds of test, PD patient walking throughout test, (d) buzzer recorded in first few seconds of test, healthy individual walking throughout test, (e) buzzer recorded in the first few seconds of test, PD patient deviates from test protocol before adhering to test instructions by starting to walk, near the end of the test, PD patient deviates from test protocol, (f) buzzer captured in the first few seconds of test, healthy individual then begins walking, after which individual deviates from test protocol and resumes walking near end of test, (g-h) PD patient and healthy individual both deviate from test instructions throughout the test.} 
\label{fig.Gait-tests-illustrative} 
\end{figure}

We present some illustrative examples of applying our hand-labelling
protocol to walking clinimetric test sensor data collected from individuals
with PD and healthy individuals. Figure \ref{fig.Gait-tests-illustrative}(a-b)
shows examples in which the user is adhering to the test protocol,
i.e. the user is apparently walking throughout the entire duration
of the test as instructed. By contrast, in Figure \ref{fig.Gait-tests-illustrative}(c-d)
the user is also walking, however, the smartphone buzzer, which is
active for a duration of approximately $2$ seconds at the start of
the test, is included in the recording.

Figure \ref{fig.Gait-tests-illustrative}(e-f) shows examples in which
the user deviates from the test protocol near the end of the test
(e), and midway through the test (f). The noticeable gap in walking
for the PD patient in test (e) may be due to ``freezing of gait''
(an absence of movement despite the intention to walk), an important
PD symptom. Nonetheless, we favour removing such instances from the
data as without additional information we cannot be sure of this identification.

\begin{figure}[htbp]

\vspace{0.2cm}
\includegraphics[width=1\columnwidth]{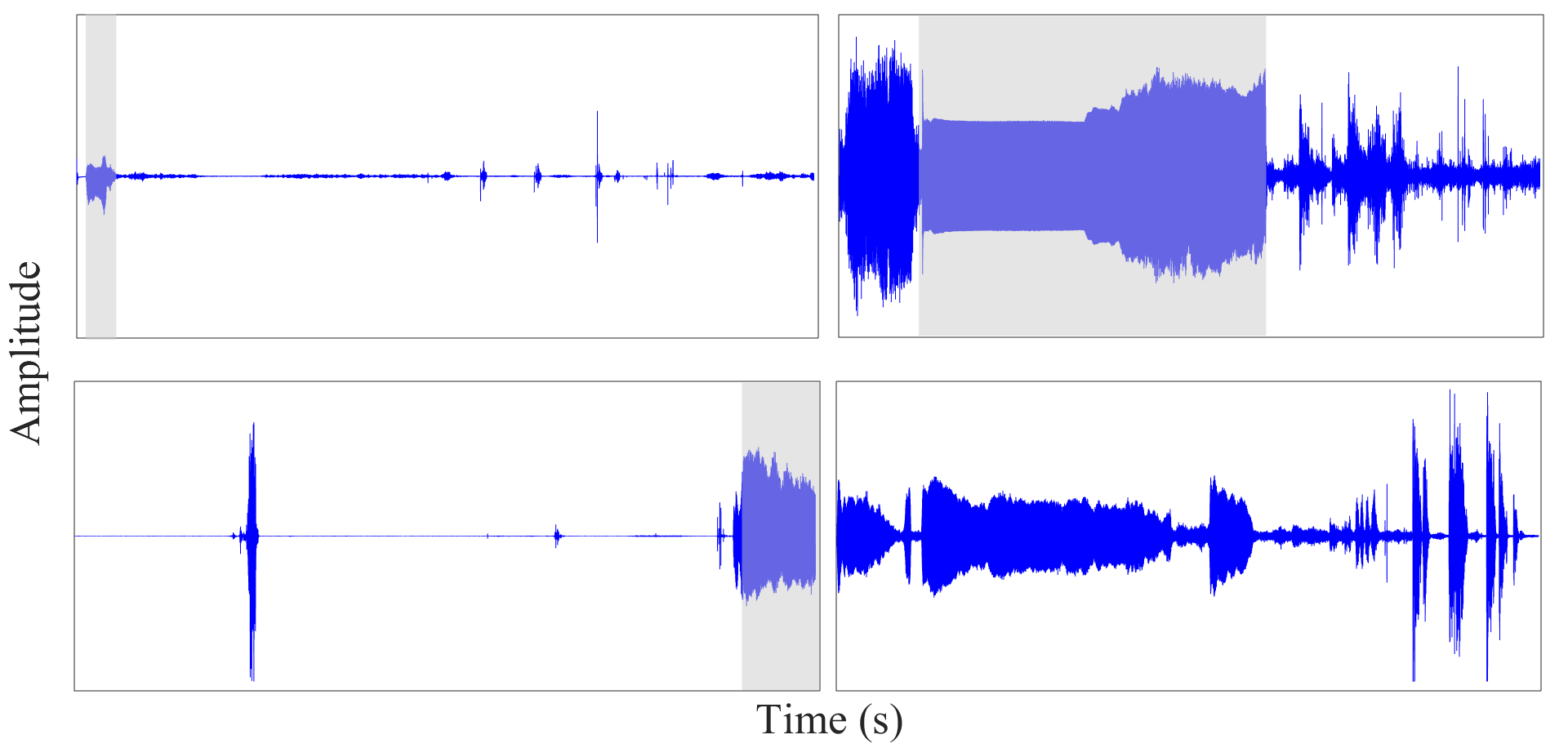}

\medskip 
\small
\caption{Illustrative examples of voice clinimetric test sensor data recorded using a smartphone microphone from healthy individuals and those with Parkinson's disease. The top two tests are performed by healthy individuals and the bottom two by PD patients. Voice tests take approximately 20s and the shaded area marks the part of the test where the user adheres to the test protocol.} 
\label{fig.Voice-tests-illustrative} 
\end{figure}

Figure \ref{fig.Gait-tests-illustrative}(g-h) shows examples in which
the user is not adhering to the test protocol throughout the duration
of the test. Such instances can occur, for example, when the user
is attempting the test for the first time, or when the user is interrupted
at the start of the test due to some unknown distraction. Similar
illustration of the voice tests can be found in Figure \ref{fig.Voice-tests-illustrative}.

\smallskip{}

\section{Sensor-specific preprocessing\label{sec:Pre-processing-of-the sensor data}}

Whenever we analyze data from sensors it is often necessary to apply
some sensor-specific processing to the raw data to remove various
confounds. This is the case for the walking, balance and voice tests
described above.

\subsection{Isolating and removing orientation changes from accelerometry data\label{subsec:Isolating-and-removing}}

\noindent One of the primary functions for which MEMs accelerometers
were included in smartphones is to detect the orientation in which
the us er is holding the device and allow for appropriate shift of
the display between ``landscape'' ( horizontal) and ``portrait'' (vertical)
display modes. The accelerometer does this by measuring the earth's
gravitational field acting on the smartphone. In recent years there
has been an increasing interest in utilizing built-in accelerometer
sensors to infer various motion patterns of the user \citep{reilly2013mobile}.
In clinimetric testing they could be used to assess the ability of
the user to perform certain daily activities which can be a strong
indicator of a particular health condition. For example, it has been
shown that PD can significantly affect activities such as gait or
standing upright. In order to use the accelerometer data collected
from a smartphone for monitoring gait (or balance), we first need
to remove the effect of the earth's gravitational field from the raw
accelerometer data as it is a confounding factor. This is because
the accelerometer output is sensitive to the orientation of the device
with respect to the gravitational field, as well as the accelerations
due to the user's motion patterns that we seek to measure. 

Let us denote the raw accelerometer output which reflects the total
acceleration due to forces applied to the device by $a_{r}\in\mathbb{R}^{3}$,
then we can write:
\begin{equation}
a_{r}=a_{d}+a_{g}
\end{equation}
where $a_{g}\in\mathbb{R}^{3}$ is the gravitational acceleration
acting on the device and $a_{d}$ is the sum of the residual accelerations
acting on the device (often called ``linear'' or ``dynamic'') acceleration.
We are interested in estimating $a_{d}$ from $a_{r}$ without observing
$a_{g}$ directly. A widely-used approach involves ``sensor fusion''
using gyroscopes or other sensors to jointly infer device orientation
\citep{Kavanagh2008}, but this approach relies on access to additional
synchronized sensor data. Without additional information we have to
make fairly strong assumptions about $a_{g}$ and $a_{d}$ in order
to infer them. A common assumption is that orientation is locally
stationary in time, so that passing the raw data through a digital
high pass filter of some sort is (under certain mathematical assumptions)
the optimal solution. However, this is too restrictive a set of assumptions
to make when the user is constantly interacting with the device and
performing activity tests at the same time. At the same time it is
reasonable to assume that the measured gravitational field will follow
relatively simple dynamics compared to the dynamic component. In this
work we propose a novel approach which models the gravitational field
as a piecewise linear signal. This assumption is less restrictive
than standard stationarity assumptions, but still allows us to rapidly
filter away the effect of device orientation.

To estimate the unknown piecewise linear trend $a_{g}$ from the raw
output $a_{r}$, we use \textit{$L_{1}$-trend filtering}, which is
a variation of the widely-used Hodrick\textendash Prescott (H-P) filter
\citep{Hodrick1997}. The $L_{1}$-trend filter substitutes a sum
of absolute values (i.e., an $L_{1}$ norm) for the sum of squares
used in H-P filtering to penalize variations in the estimated trend. 

\begin{figure}[htbp]  

\includegraphics[width=1\columnwidth]{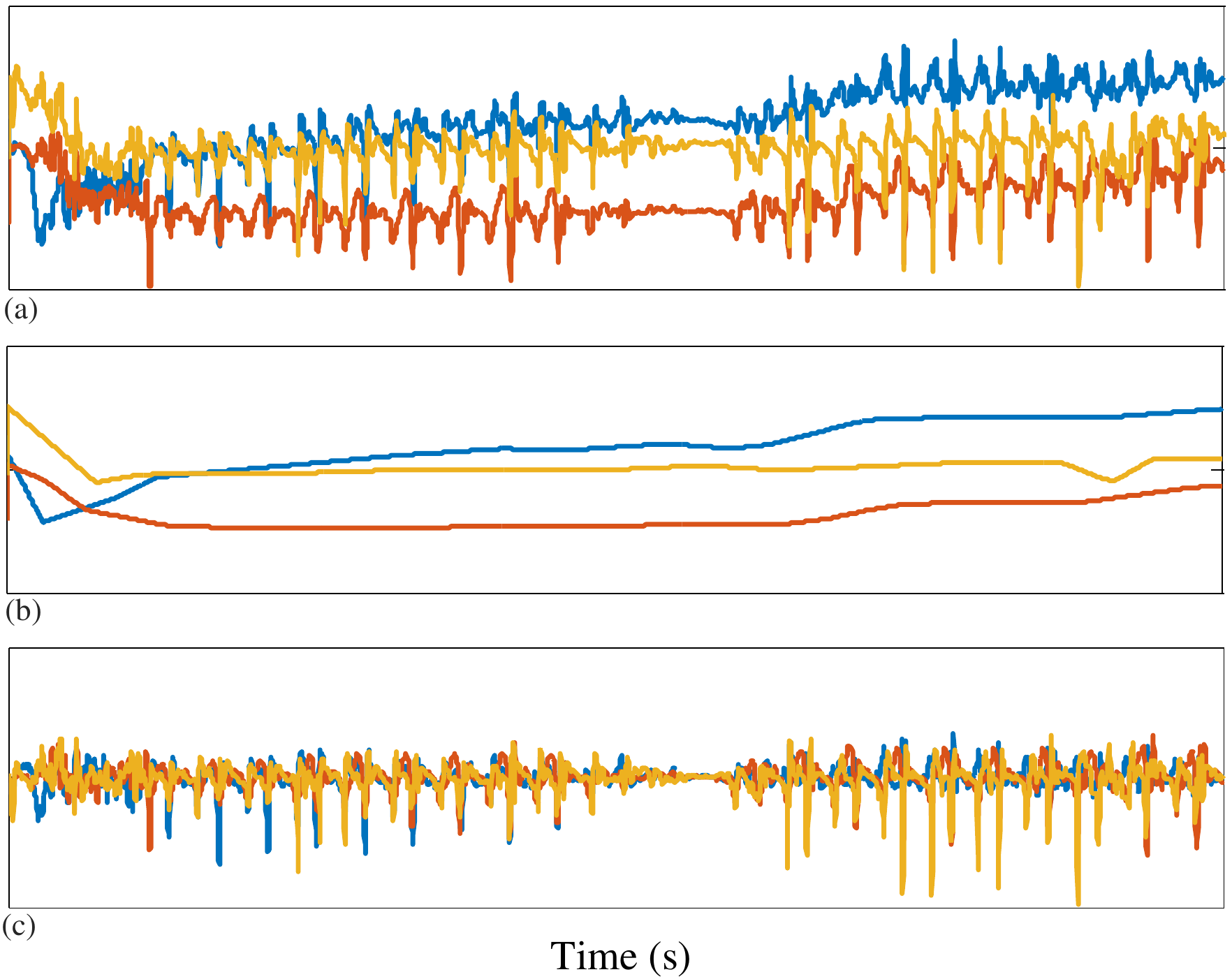}

\caption{Raw accelerometer sensor output vector $x$ for a single walking test (top panel); (a) raw acceleration data, (b) estimated gravitational orientation trend $\hat{g}$, (c) estimated dynamic acceleration after removing the effect of device orientation.} 
\label{fig.GravityGoodAxesOff} 
\end{figure}

Assume we have $T$ measurements of the raw accelerometer data ($T$
data points) and let us denote them by $x_{1},\dots,x_{T}$ where
$x_{t}\in\mathbb{R}^{3}$ for $t=1,\dots,T$. The trend vectors $g_{1},\dots,g_{T}$
should minimize the objective function: 
\begin{equation}
\hat{g}=\textrm{arg}\min_{g}\left\{ \frac{1}{2}\sum_{t=1}^{T}\left|x_{t}-g_{t}\right|+\lambda\sum_{t=2}^{T-1}\left|g_{t-1}-2g_{t}+g_{t+1}\right|\right\} \label{eq:L1-Trend Objective}
\end{equation}
where $\hat{g}=\hat{g}_{1},\dots,\hat{g}_{T}$ denotes the set of
gravitational vectors minimizing the functional in \eqref{eq:L1-Trend Objective}.
The linear acceleration is then estimated by subtracting the estimated
gravitational trends from the raw sensor output: $x_{t}^{d}=x_{t}-\hat{g}_{t}$
for $t=1,\dots,T$, see Figure \ref{fig.GravityGoodAxesOff}.

\subsection{Feature extraction}

\noindent In contrast to existing algorithms for segmentation of sensor
data, we propose a simple approach which can use just a single feature
for each kind of sensor. The system could be easily extended to include
a more sophisticated feature engineering stage, however the benefits
of this will depend upon the clinimetric tests analyzed.

We compute the magnitude of the 3-axis dynamic acceleration vector
estimated after preprocessing to remove the gravitational orientation
component. If we denote the dynamic acceleration at time $t$ by $x_{t}\in\mathbb{R}^{3}$,
then the magnitude is the Euclidean norm $\left\Vert x_{t}\right\Vert _{2}=\sqrt{x_{t,1}^{2}+x_{t,2}^{2}+x_{t,3}^{2}}$,
and this is proportional to the magnitude of the instantaneous dynamic
force being applied to the device (the missing constant of proportionality
here is the combined, but unknown, mass of the device and the wearer).
For quality control of both the walking and balance tests we do not
directly model the dynamic acceleration vector, only its magnitude
(for the balance tests) and $\log_{10}$ magnitude (for the walking
tests).

In order to efficiently process the data from the voice test we also
extract a single feature from the raw sensor output. The raw voice
data used in this study is sampled at 44,100 Hz and direct segmentation
of this very high-rate signal would be unnecessarily computationally
challenging. Instead we segment the original signal into 10ms windows
and extract the signal energy of the data in each of those windows
which contains 441 unidimensional (and dimensionless) sensor measurements.
If we denote the microphone output as $x_{1},\dots,x_{T}$, the first
window consists of $\left\{ x_{1},\dots,x_{441}\right\} $; the second
window $\left\{ x_{442},\dots,x_{882}\right\} $ etc. The signal energy
associated with each window is the squared Euclidean norm of the measurements
in that frame with the first window denoted by $\epsilon_{1}=\sqrt{\sum_{t=1}^{441}x_{t}^{2}}$.

\subsection{Downsampling (sample rate reduction)}

\noindent While we are interested in processing the data at sufficiently
high frequency, in some situations modeling the raw directly can be
computationally wasteful when our interest is quality control only.
We have studied the power spectrum of the different tests to find
when we can downsample the original high frequency signal to a lower
frequency without losing essential information. In practice appropriate
downsampling is particularly important whenever we use AR models.
This is because high frequency data requires inferring high number
of AR coefficients to accurately capture the dynamics of the data.
Estimating large numbers of AR coefficients is difficult because parameter
inference in the model requires high computational effort, and since
the amount of data is always limited it is also more likely to lead
to unreliable estimates for the AR model parameters.

In general, obtaining a representation of a signal which is invariant
to downsampling is an ill-posed problem. However, in the special case
of \textit{bandlimited} signals it can be shown that ideal reconstruction
is possible as long as certain criteria hold. Bandlimited signals
have restricted support in the frequency domain such that their Fourier
transform is $0$ for frequencies $\omega$ for which $\left|\omega\right|>2\pi B$
where $B>0$ is the \textit{bandwidth} of the signal which reflects
the maximum frequency content (Hz). In other words the spectrum of
bandlimited signals have support bounded at $B$. Given that a signal
is bandlimited we can produce an ideal reconstruction of it as long
as we have samples from it at a frequency of at least $2B$. This
minimum sampling frequency requirement which allows for perfect reconstruction
is known as the \textit{Nyquist criterion} and specifies that the
longest sampling time duration which ensures perfect reconstruction
is $1/\left(2B\right)$. The ideal reconstruction of bandlimited signals
from a limited number of samples given that the Nyquist criterion
holds can be performed using \textit{\emph{the }}\textit{Shannon-Whittaker
reconstruction formula} \citep{mallat2008wavelet}. If we assume that
the high frequency data recorded by the sensors consists of samples
from a real signal $f\,:\mathbb{\,R}\to\mathbb{R}$, then if $f$
is bandlimited, according to the Nyquist criterion we can sample the
original data at a rate near $2B$ and reconstruct $f$ perfectly
from the downsampled data.

\begin{figure}[htbp]  

\includegraphics[width=1\columnwidth]{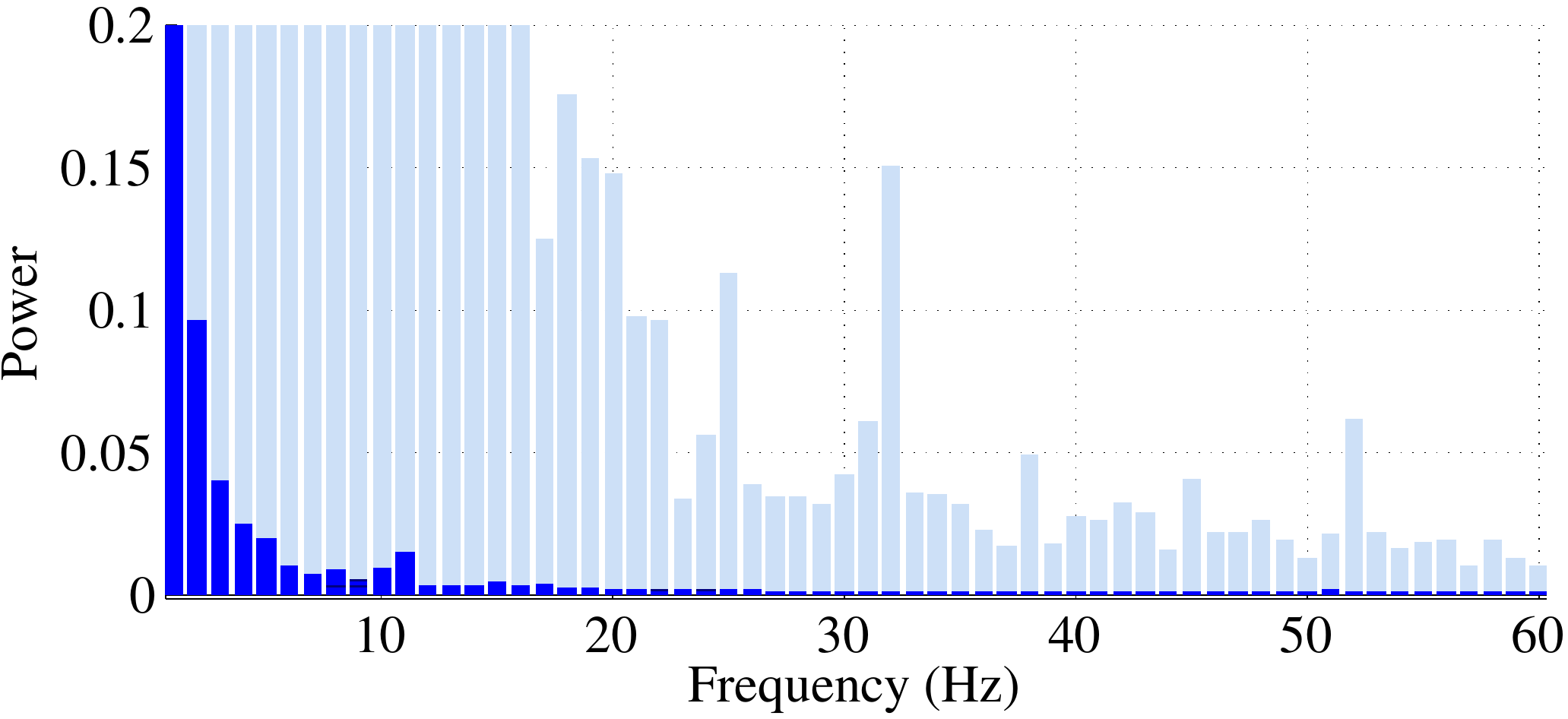}

\caption{Power spectra of the magnitude of the 3-axis accelerometer data from each of the 100 balance and walking tests. The 200 densities are plotted on top of each other in order to see the maximum support that was observed at each frequency over all 200 tests. The shaded bars display the power spectra from the balance tests and the colored bars denote the spectra of the walking tests. Since walking is highly periodic, in the spectra of these tests most of the power is found in the lower frequencies associated with periodic walking. During the balance tests periodic activities are observed for very short time durations and the only periodic activity consistently recorded is the smartphone buzzer. This explains why we see that a lot of the power in the spectra is not found at the lower frequencies, but instead spread across the higher frequencies.}
\label{fig.100BalanceAndGaitTestsCombinedPowerSpectrum} 
\end{figure}

In the real world most signals are not exactly bandlimited, but their
power spectrum shows small magnitude at high frequencies, thus we
can apply a low pass filter to the original signal to make it bandlimited.
We evaluated the power spectrum of the accelerometer data from each
of the 100 walking tests and the 100 balance tests. In Figure \ref{fig.100BalanceAndGaitTestsCombinedPowerSpectrum}
we have combined all 100 densities for the walking tests and all 100
densities for the balance tests in the same plot. Figure \ref{fig.100BalanceAndGaitTestsCombinedPowerSpectrum}
suggest strong evidence that the $\log_{10}$ magnitude of the linear
acceleration from the walking tests comes from a nearly bandlimited
signal with bandlimit $B<15\text{Hz}$. Therefore, after removing
the effect of the gravitational component all the data coming from
the walking tests is preprocessed with a low-pass filter of a cut-off
frequency of 15Hz. Following the Nyquist criterion then we can downsample
the signal to uniform sampling rate of $2\times15=30\text{Hz}$. 

There is little evidence (see Figure \ref{fig.100BalanceAndGaitTestsCombinedPowerSpectrum})
to support the interpretation that the sensor data for the balance
tests is bandlimited, therefore we omit the downsampling step and
model the magnitude of the accelerometer data from the balance tests
in its original, high sample-rate form.

\section{Sequential behaviour modelling\label{sec:Sequential-behaviour-modeling}}

\noindent It is realistic to assume that for most remote health monitoring
technology, detailed user behaviour information would never be available
after deployment. Therefore, traditional supervised machine learning
activity recognition systems are not applicable and we turn to unsupervised
learning. We propose two different methods for segmenting distinct
behaviours.

The first method is based on fitting a GMM to the data generated from
each clinimetric test. The method does not require any labeled data
for training, but imposes the strong assumption that the data violating
the test protocols can be clustered into a different Gaussian component
to data which adheres to the protocol. Despite the simplicity of this
method, we demonstrate that in some scenarios it manages to segment
out most of the bad quality data points with very little computation
involved.

For more complex scenarios we also propose a general technique which
can be used to segment different behaviours based on the properties
of the data into some estimated number of different ``states''.

\begin{figure*}[htbp]  

\includegraphics[width=2.05\columnwidth]{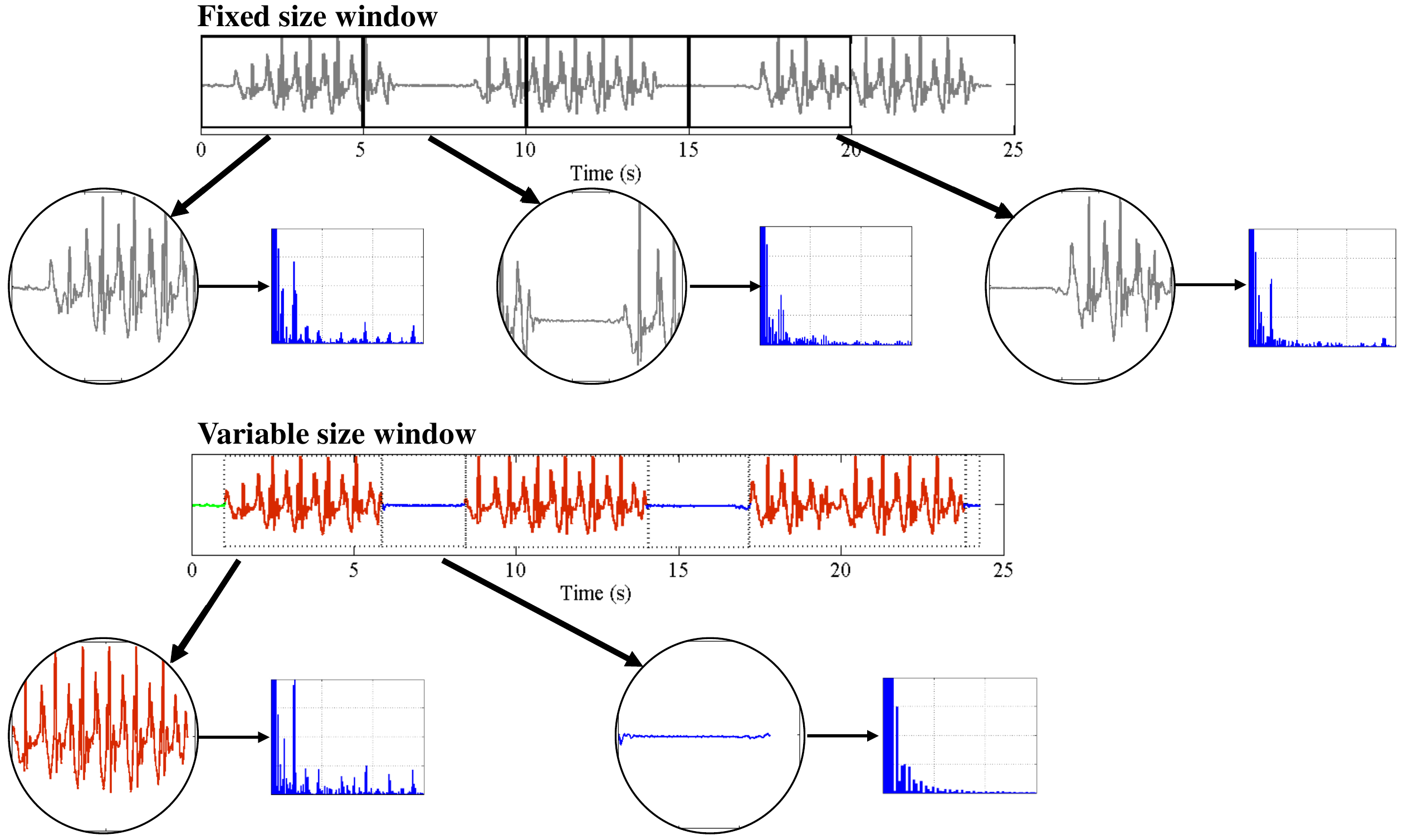}

\caption{Demonstration of how the traditional feature extraction approach relying on windowing with fixed window size can lead to misleading estimates of any features from the frequency domain. The bottom plot demonstrates how an accelerometer signal is segmented using a switching AR model and more accurate estimates of frequency domain features can be obtained as a result. We observe this by looking at how the power spectra changes.}
\label{fig.EffectOfWindowing} 
\end{figure*}

\subsection{Unsupervised behaviour modeling}

\noindent One of the primary methodological contributions of this
work is obviating the need for elaborate feature engineering. Existing
machine learning methods for activity recognition and segmentation
of sensor data typically involve windowing the data at the start of
the analysis and extracting a rich set of features from each data
window with prespecified fixed length (such as 30s, 1min etc.). When
it comes to the problem of quality control, in particular for clinimetric
data, there are two major problems with this existing approach:
\begin{enumerate}
\item Any behavior changes which occur in the data within a window cannot
be represented (see Figure \ref{fig.EffectOfWindowing}). Due to our
inability to model and account for them, they confound the feature
values for the window in which they occur. Many of the features used
for processing sensor data are some type of frequency domain feature
(for example, dominant frequency component; largest magnitude Fourier
coefficients; various wavelet coefficients etc.). Frequency domain
features are only meaningful for signals that have no abrupt changes;
Fourier analysis over windows which contain abrupt discontinuities
is dominated by unavoidable Gibb's phenomena \citep{little2011generalized}.
Unfortunately, behavioural data from clinimetric tests is rife with
such discontinuities due to inevitable changes in activities during
tests. In the approach we propose, the window sizes and boundaries
adapt to the data, since segmentation is learned using a probabilistic
model which is specifically designed to capture rapid changes in activity
when they occur, but also to model the intricacies of each activity.
\item The optimal features to be extracted from each window depend largely
on the task/activity occurring in that window. If we are interested
in developing a unified framework which works under a realistically
wide set of scenarios encountered outside the lab, hand-picking an
appropriate set of features for each activity that a clinimetric test
might include is not feasible. This issue could be partially overcome
if we use ``automated'' features such as principle component analysis
(PCA), but this entails unrealistic assumptions (i.e. linearity).
Alternatively, we could use an unsupervised approach for automated
feature learning such as layers of restricted Boltzmann machines (RBM)
or deep belief networks (DBN). However, these methods require large
volumes of data from every behaviour (which is unlikely to ever be
available from health-impaired users), and sufficient computational
power to train, making them unsuitable for deployment in real-time
applications on smartphones or other resource-constrained devices.
Even so, basic RBMs and DBNs would still need to be trained on features
extracted after windowing of the sensor data. Additionally, although
the features extracted using deep learning systems have demonstrated
highly accurate classification results for many applications, we lack
any ability to interpret these models to give a human understanding
of what aspects of the data they represent. When dealing with healthcare
applications, this lack of interpretability could significantly reduce
the explanatory power required to gain confidence in the technique.
The system we propose does not rely on extensive feature engineering
or inscrutable deep learning algorithms, since we demonstrate sufficiently
high performance using a single feature for each of the different
sensor types. Of course, the proposed approach can be easily extended
to use multiple features per data type and this could potentially
boost performance when appropriate features are chosen.
\end{enumerate}

\paragraph{Segmentation with GMMs\label{par:Segmentation-with-GMMs}}

\begin{figure}[htbp]  
\subfigure[]{\includegraphics[width=0.45\columnwidth]{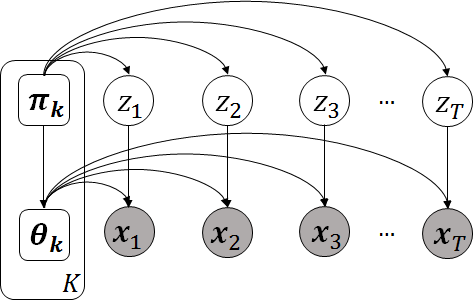}}
\hfill
\subfigure[]{\includegraphics[width=0.55\columnwidth]{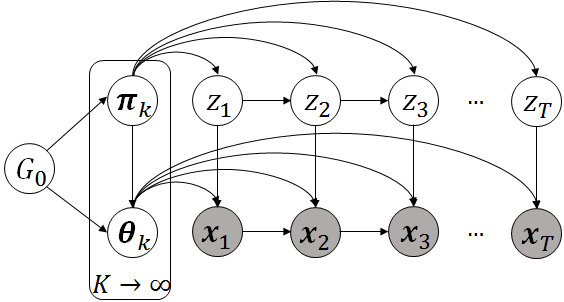}}
\caption{Probabilistic graphical models for (a) Gaussian mixture model (GMM) and (b) nonparametric switching autoregressive (AR) model. The GMM component parameters $\theta_k$ are the mean and the variance of data associated with component $k$. By contrast, the component parameters $\theta_k$ for the switching AR consist of the AR coefficients $A_1^k,\dots,A_r^k$ and the AR error parameters $\mu_k$ and $\sigma_k$. Parameter $\pi_k$ denotes the mixing coefficients and the transition matrix for the GMM and the switching AR, respectively. In the parametric GMM $\pi_k$ and $\theta_k$ are fixed. In the nonparametric switching AR $\pi_k$ and $\theta_k$ are modeled with an HDP prior, where  $G=\{G_1,\dots,G_{K^+}\}\sim$HDP$\left(\alpha,\gamma,\theta_0\right)$. Hyperparameter $\theta_0$ denotes the conjugate prior over the $A's$, the $\mu's$ and the $\sigma's$. See main text for further description of the HDP and its concentration parameters $\alpha$ and $\gamma$.}
\label{fig.SamplesLDS} 
\end{figure}

\noindent The simpler proposal is a GMM-based approach which relies
on the assumption that (at least most of the time) the magnitude of
the sensor data adhering to the test protocols is different from the
magnitude of the data violating them; or that we can approximately
cluster the magnitude of that data into two separate Gaussian components.
After we have applied the appropriate pre-processing depending on
the data source (as described above) we fit a two-component ($K=2$)
GMM to each of the different sensor datasets. The GMMs are estimated
in an unsupervised way using the expectation-maximization (E-M) algorithm
where after convergence points are clustered to their most likely
component using the maximum-a-posteriori (MAP) principle \footnote{More precisely, each point in time is assigned to the component that
maximizes the probability of its component indicator.}. Let us denote the preprocessed data by $x_{1},\dots,x_{T}$ with
$T$ being the number of sensor outputs for a given test after pre-processing.
By fitting a $K=2$ component GMM to $x_{1},\dots,x_{T}$ we will
estimate some \emph{indicators} $z_{1},\dots,z_{T}$ which denote
the component assignment of each time point (for example $z_{t}=1$
denotes that time point $x_{t}$ is associate with component 1). We
denote by $\left(\mu_{1},\sigma_{1}\right)$ and $\left(\mu_{2},\sigma_{2}\right)$
the component mean and variance for the first and second component
respectively. We use the estimated means $\mu_{1}$ and $\mu_{2}$
to identify whether the component corresponds to protocol adherence
or violation. For walking tests and voice tests we assume that if
$\mu_{1}>\mu_{2}$ all data points $\left\{ x_{t}:\,z_{t}=1\right\} $
represent adherence to the test protocols, hence $\left\{ x_{t}:\,z_{t}=2\right\} $
represent protocol violation. By contrast, for the balance tests we
assume that time points associated with the larger mean represent
violation and the points associated with the smaller mean represent
adherence to the protocols. Since the GMM ignores the sequential nature
of the data (see GMM graphical model, Figure \ref{fig.SamplesLDS}),
the estimated indicators $z_{1},\dots,z_{T}$ can switch very rapidly
between the two components, providing an unrealistic representation
of human behaviour. In order to partially address this issue, we apply
moving \textit{median filtering} \citep{arce2005nonlinear} to the
indicator $z_{1},\dots,z_{T}$ and run it repeatedly to convergence.
In this way we obtain a ``smoothed'' sequence $\hat{u}_{1},\dots,\hat{u}_{T}$
which we use as classification of whether each of $x_{1},\dots,x_{T}$
is adhering to, or violating, the relevant protocol; time point $t$
is classified as adherence if $\hat{u}_{t}=1$ and violation if $\hat{u}_{t}=2$.

\paragraph{Segmentation with the switching AR model}

\noindent In order to extend the GMM to model long time-scale dependence
in the data we can turn to \emph{hidden Markov models} (HMM) \citet{Rabiner1989}.
HMMs with Gaussian observations (or mixtures of Gaussian observations)
have long dominated areas such as activity \citep{toreyin2008flame,andrade2006hidden,chung2008daily,gao2006behavioral,Raykov2016}
and speech recognition \citep{Rabiner1989,juang1991hidden,gales2008application}.
However, simple HMMs fail to model any of the frequency domain features
of the data, and are therefore not flexible enough to describe the
sensor data; instead we need to use a more appropriate model.

The \textit{switching autoregressive} model is a flexible discrete
latent variable model for sequential data which has been widely used
in many applications, including econometrics and signal processing
\citep{oh2008learning,jilkov2004online,li2007efficient,chiang2008hidden}.
Typically some $K$ number of different AR models are assumed a-priori.
An order $r$ AR model is a random process which describes a sequence
$x_{t}$ as a linear combination of previous values in the sequence
and a stochastic term:
\begin{equation}
x_{t}=\sum_{j=1}^{r}A_{j}x_{t-j}+e_{t}\qquad e_{t}\sim\mathcal{N}\left(0,\sigma^{2}\right)\label{eq:Autoregressive process}
\end{equation}
where $A_{1},\dots,A_{r}$ are the AR coefficients and $e_{t}$ is
a zero mean, Gaussian i.i.d. sequence (we can trivially extend the
model such that $e_{t}\sim\mathcal{N}\left(\mu,\sigma^{2}\right)$
for any real-valued $\mu$). An important property of AR models is
that we can express its \emph{power spectral density} as a function
of its coefficients:
\begin{equation}
S\left(f\right)=\frac{\sigma^{2}}{\left|1-\sum_{j=1}^{r}A_{j}\exp\left(-i2\pi fj\right)\right|^{2}}
\end{equation}
where $f\in\left[-\pi,\pi\right]$ is the frequency variable with
$i$ here denoting the imaginary unit. This means that the order of
the AR model directly determines the number of ``spikes'' in its spectral
density, which corresponds to the number of zeros in the numerator
of this expression, and therefore the complexity or amount of detail
in the power spectrum of $x_{t}$ that can be represented.

In switching AR models we assume that the data is an inhomogeneous
stochastic process and multiple different AR models are required to
represent the dynamic structure of the series, i.e.:
\begin{equation}
x_{t}=\sum_{j=1}^{r}A_{j}^{z_{t}}x_{t-j}+e_{t}^{z_{t}}\qquad e_{t}^{z_{t}}\sim\mathcal{N}\left(0,\sigma_{z_{t}}^{2}\right)
\end{equation}
where $z_{t}\in\left\{ 1,\dots,K\right\} $ indicates the AR model
associated with point $t$. The latent variables $z_{1},\dots,z_{T}$
describing the switching process are modeled with a Markov chain.
Typically, $K\ll T$ allowing us to cluster together data which is
likely to be modeled with the same AR coefficients.

The switching AR model above is closely related to the HMM: as with
the switching AR model, the HMM also assumes that data is associated
with a sequence of hidden (latent) variables which follow a Markov
process ($z_{1},\dots,z_{T}$ in Figure \ref{fig.SamplesLDS}). However,
in the case of HMMs we assume that given the latent variables the
observed data is independent. In other words, the simplest HMM can
be considered as a switching AR model where the order $r$ of each
AR is $0$ with non-zero mean error term. Neither of the models discussed
here are necessarily limited to Gaussian data and there have been
HMM extensions utilizing: multinomial states for part-of-speech tagging
\citep{Goldwater2007fully}, Laplace distributed states for passive
infrared signals \citep{Raykov2016} or even neural network observational
models for image and video processing \citep{johnson2016composing}.

The segmentation produced with any variant of the HMM is highly dependent
on the choice of $K$ (the number of hidden Markov states i.e. distinct
AR models). In the problem we study here, the number $K$ would roughly
correspond to the number of different behavioural patterns which occur
during each of the clinimetric tests. However, it is not realistic
to assume we can anticipate how many different behaviours can occur
during each test. In fact it is likely that as we collect data from
more tests, new patterns will emerge and $K$ will need to change.
This motivates us to seek a \emph{Bayesian nonparametric} (BNP) approach
to this segmentation problem: a BNP extension of the switching AR
model described above which will be able to accommodate an unknown
and changing number $K^{+}$ of AR models.

The nonparametric switching AR model (first derived as a special case
of nonparametric switching linear dynamical systems in \citet{Fox2009})
is obtained by augmenting the transition matrix of the HMM underlying
the switching AR with a \emph{hierarchical Dirichlet process} (HDP)\citep{Teh2004}
prior (see Figure \ref{fig.SamplesLDS}). Effectively, the HMM component
of the switching AR model is replaced with an \emph{infinite} HMM
\citep{Beal2002}. The infinite HMM avoids fixing the number of states
$K$ in the Markov model; instead it assumes that the number of HMM
states an unknown, and potentially large $K^{+}$, and depends upon
the amount of training data we have already seen. Whenever we are
fitting an infinite HMM we typically start by assigning the data into
a single hidden state (or a small fixed number of states) and at each
step with some probability we increase the number of effective states
at each inference pass through the signal. In this way it is possible
to infer the number of effective states in an infinite HMM as a random
variable from the data. The parameters specifying how quickly the
number of effective states grows are called \emph{local }and \emph{global
concentration }hyperparameters: $\alpha$ denotes the local and $\gamma$
the global concentration.

The local $\alpha$ controls how likely it is that new types of transitions
occur between the effective states, or essentially how sparse is the
HMM transition matrix. The global $\gamma$ reflects how likely is
it for a new effective state to arise, or how many rows the transition
matrix has. Unlike the fixed $K$ in standard parametric HMMs, the
hyperparameters $\alpha$ and $\gamma$ of the infinite HMM (or any
of its extensions) can be tuned with standard model selection tools
which compute how the value of the \emph{complete data likelihood
}changes as $\alpha$ and $\gamma$ change. This allows us to model
the behavioural patterns in the smartphone clinimetric tests in a
completely unsupervised way. For a lengthier discussion and derivation
of the infinite HMM and the nonparametric switching AR model, we refer
readers to \citep{Beal2002,Teh2004,Fox2009}.

As \citet{Fox2009} remarks, the switching AR model can be also be
obtained as a special case of a more general switching\textit{ }\textit{\emph{linear
dynamical systems}}\emph{ }(LDS) model \citep{Aoki1991}. In the LDS
we assume that the observations are noisy measurements of the quantity
of interest. The switching LDS allows for even more flexible segmentation
analysis of the different behavioural states, but this comes at the
cost of substantial increase in model complexity. Additional analysis
showed that for the chosen data, the most parsimonious solution is
offered by the switching AR.

\subsection{Segmentation context mapping\label{subsec:Segmentation-context-mapping}}

\noindent The switching AR model groups together intervals of the
preprocessed data that have similar dynamics described by the same
AR pattern, i.e. we group points $x_{t}$ according to their corresponding
indicator values $z_{t}=k$ for $k\in\left\{ 1,\dots,K^{+}\right\} $.
The generality of this principle allows us to apply the framework
widely across different data sets generated from diverse clinimetric
tests such as walking, balance or voice tests.

A trained expert can reasonably identify intervals of walking or balancing
that adhere to the corresponding test protocols, while specific physical
activities would be difficult to identify purely from the accelerometer
output. Lack of behaviour labels can challenge our understanding of
the segmentation from the previous stage. This motivated the collection
of additional controlled clinimetric tests to shed some light on the
patterns we discover using the nonparametric switching AR model. The
controlled smartphone tests have been performed by healthy controls.
We collect 32 walking, 32 balance and 32 voice tests in which we vary
the orientation and location of the phone during a simulated clinimetric
test. During these tests the participants are instructed to perform
some of the most common behaviours which we observe during clinimetric
tests performed outside the lab. Activities conducted during the tests
include, freezing of gait, walking, coughing, sustained phonation,
keeping balance, and several others. A human expert annotates the
monitored behaviours with $b_{1},\dots,b_{T}$ which associate each
data point with a behavioural label (i.e. $b_{t}=\text{"walking"}$
means point $x_{t}$ was recorded during walking). By contrast to
the clinimetric tests performed outside the lab, here we have relatively
detailed information about what physical behaviour was recorded in
each segment of these controlled tests.

Since we have the ``ground truth'' labels $b$ for the controlled
clinimetric tests, we can be confident in the interpretation of the
intervals estimated by the unsupervised learning approach. This allows
us to better understand the different intervals inferred from data
collected from outside the lab, when labels $b$ are not available.
Note that $b_{1},\dots,b_{T}$ are not used during the training of
the nonparametric switching AR model, but only for validation. Furthermore,
the distribution of the data from the actual clinimetric tests collected
outside of the lab significantly departs from the distribution of
the data of the controlled tests.

We assess the ability of the model to segment data consisting of different
behaviours. This is done by associating each of the unique $K^{+}$
values that the indicators $z$ can take with one of the behavioural
labels occurring during a controlled test. For each $k\in\left\{ 1,\dots,K^{+}\right\} $,
state $k$ is assumed to model behaviour $b_{k}$ with $b_{k}=\text{mode}\left\{ b_{t}:z_{t}=k\right\} $
being the most probable behaviour during that state.

Using this simple mapping from the numerical indicators $z_{1},\dots,z_{T}$
to interpretable behaviours, we obtain estimated behaviour indicators
$\hat{z}_{1},\dots,\hat{z}_{T}$. Using the estimated behaviour indicators
$\hat{z}_{1},\dots,\hat{z}_{T}$ and the ``ground truth'' labels $b_{1},\dots,b_{T}$
we compute the following algorithm performance measures: balanced
accuracy (BA), true positive (TP) and true negative (TN) rates for
the segmentation approach in Table \ref{tab:Behaviour recognition accuracy}.
For example given behaviour $b^{*}$, these metrics are computed using:

\begin{equation}
\begin{array}{c}
\text{TP}=\frac{\sum_{t=1}^{T}\mathbf{1}\left(\hat{z}_{t}=b^{*}\cap b_{t}=b^{*}\right)}{\sum_{t=1}^{T}\mathbf{1}\left(\hat{z}_{t}=b^{*}\right)};\text{TN}=\frac{\sum_{t=1}^{T}\mathbf{1}\left(\hat{z}_{t}\neq b^{*}\cap b_{t}\neq b^{*}\right)}{\sum_{t=1}^{T}\mathbf{1}\left(\hat{z}_{t}\neq b^{*}\right)};\\
\\
\textrm{BA}=\frac{\text{TP}+\text{TN}}{2}
\end{array}\label{eq:Accuracy metrics}
\end{equation}
where $\mathbf{1}\left(\cdot\right)$ denotes the indicator function
which is $1$ if the logical condition is true, zero otherwise.

\begin{table}
\caption{\label{tab:Behaviour recognition accuracy}}
Balanced accuracy (BA), true positive (TP) and true negative (TN)
rates for the nonparametric switching AR model trained on walking
and balance tests performed in a controlled environment. The TP rate
reflects the ability to correctly identify an activity when occurring
and the TN rate reflects the ability to correctly indicate lack of
that activity whenever not occurring. \smallskip{}
\par
\centering{}%
\begin{tabular}{|c|c|c|c|}
\hline 
Behaviour & BA & TP & TN\tabularnewline
\hline 
\hline 
Walking & 95\% & 96\% & 93\%\tabularnewline
\hline 
Standing up straight & 95\% & 98\% & 91\%\tabularnewline
\hline 
Phone stationary & 98\% & 100\% & 95\%\tabularnewline
\hline 
Sustained phonation & 98\% & 99\% & 97\%\tabularnewline
\hline 
\end{tabular}
\end{table}

Outside the lab we cannot always label physical behaviours with high
confidence. Instead, we use binary labels $u_{1},\dots,u_{T}$ which
take values $u_{t}=1$ if point $x_{t}$ adheres or $u_{1}=2$ if
it violates the applicable test protocol (as described in Section
\ref{subsec:Hand-labelling-for-algorithm}). In order to classify
a time point $x_{t}$ with respect to its adherence to the protocol,
it is sufficient to simply classify the state assignment $z_{t}$
associated with that time point.

To automate this context mapping, we use a very highly interpretable
\emph{naive Bayes classifier}. We train the classifier using the posterior
probabilities\footnote{We noticed that we can obtain very similar accuracy using just the
modal estimates of the indicators $z_{1},\dots,z_{T}$ as an input
to the classifier, which takes substantially less computational effort
compared to computing the full posterior distribution of the indicators.} of the indicators $z_{1},\dots,z_{T}$ associated with the training
data as inputs and the corresponding binary labels $u_{1},\dots,u_{T}$
as outputs. For a new test point $\tilde{x}$ we can then compute
the vector of probabilities $P\left(\tilde{z}\left|x_{1},\dots,x_{T},\theta,\pi\right.\right)$
given the switching AR parameters $\theta$ and $\pi$ (explained
in Figure \ref{fig.SamplesLDS}) and rescale them appropriately to
appear as integer frequencies; we will write those vectors of frequencies
as $p_{\tilde{z}}=\left(p_{\tilde{z},1},\dots,p_{\tilde{z},K^{+}}\right)$.
The naive Bayes classifier assumes the following probabilistic model:
\begin{equation}
P\left(p_{\tilde{z}}\left|u_{1},\ldots,u_{T},\tilde{u},p_{z_{1},\ldots,z_{T}}\right.\right)=\frac{\left(\sum_{k=1}^{K^{+}}p_{k}\right)!}{\prod_{k=1}^{K^{+}}p_{k}!}\prod_{k=1}^{K^{+}}\bar{\pi}_{k,\tilde{u}}^{p_{k}}\label{eq:naive Bayes rule}
\end{equation}
where $\bar{\pi}_{k,\tilde{u}}$ denotes the training probability
for attribute $k$ given observation is from class $\tilde{u}$. This
model can be then reversed (via Bayes rule) to predict the class assignment
$\hat{u}\in\left\{ 1,2\right\} $, for some unlabeled input $p_{\tilde{z}}$:
\begin{equation}
\hat{u}=\argmax_{c\in\left\{ 1,2\right\} }\left[\log P\left(\tilde{u}=c\right)+\sum_{k=1}^{K+}p_{k}\log\left(\bar{\pi}_{k,c}\right)\right]
\end{equation}
with $P\left(\tilde{p}=c\right)$ enabling control over the prior
probabilities for class adherence/violation of the protocols.

\begin{figure}[htbp]  

\includegraphics[width=1.01\columnwidth]{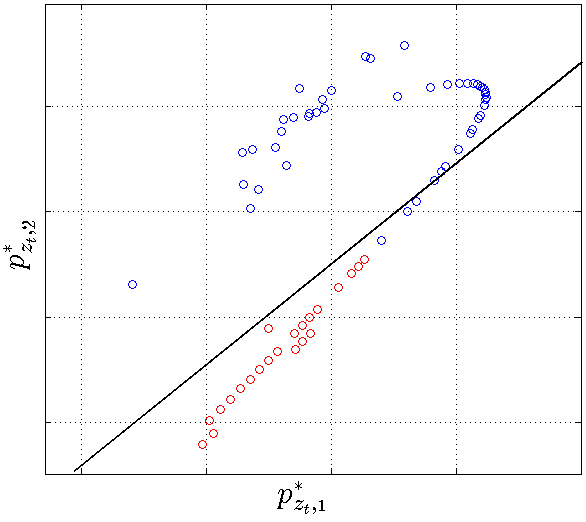}

\caption{Visualizing the context mapping of behavioural segmentation to clinimetric protocol adhere/violation predictions, data from a single walking test. The log of the input segmentation state probabilities $p_{z_1},\dots,p_{z_T}$ is projected into 2-D using linear discriminant analysis (LDA), projections are denoted with $p^*_{z_t}$ where the $p_{z_t}$ is a $K^+$ dimensional vector and $p^*_{z_t}$ is 2-D. Input projections are labeled with adherence (red) and violation (blue). The ``linear'' structure can be explained with the fact that typically, a data point $x_t$ is associated with high probability to only 1 of $K^+$ AR states in the behavioural segmentation (and low probability for the remaining AR states) so that the input vector $p_{z_t}$ is sparse. The decision boundary of the multinomial naive Bayes classifier is also projected (using the same LDA coefficients) onto 2-D (black line). A few outlier projections $p^*_{z_t}$ are outside the plot axis limits, but they do not affect significantly the decision boundary and yet reduce the visual interpretability of the projection of the bulk of the data.} 
\label{fig.LDAprojection}
\end{figure}

\begin{figure}[htbp]  

\includegraphics[width=1.01\columnwidth]{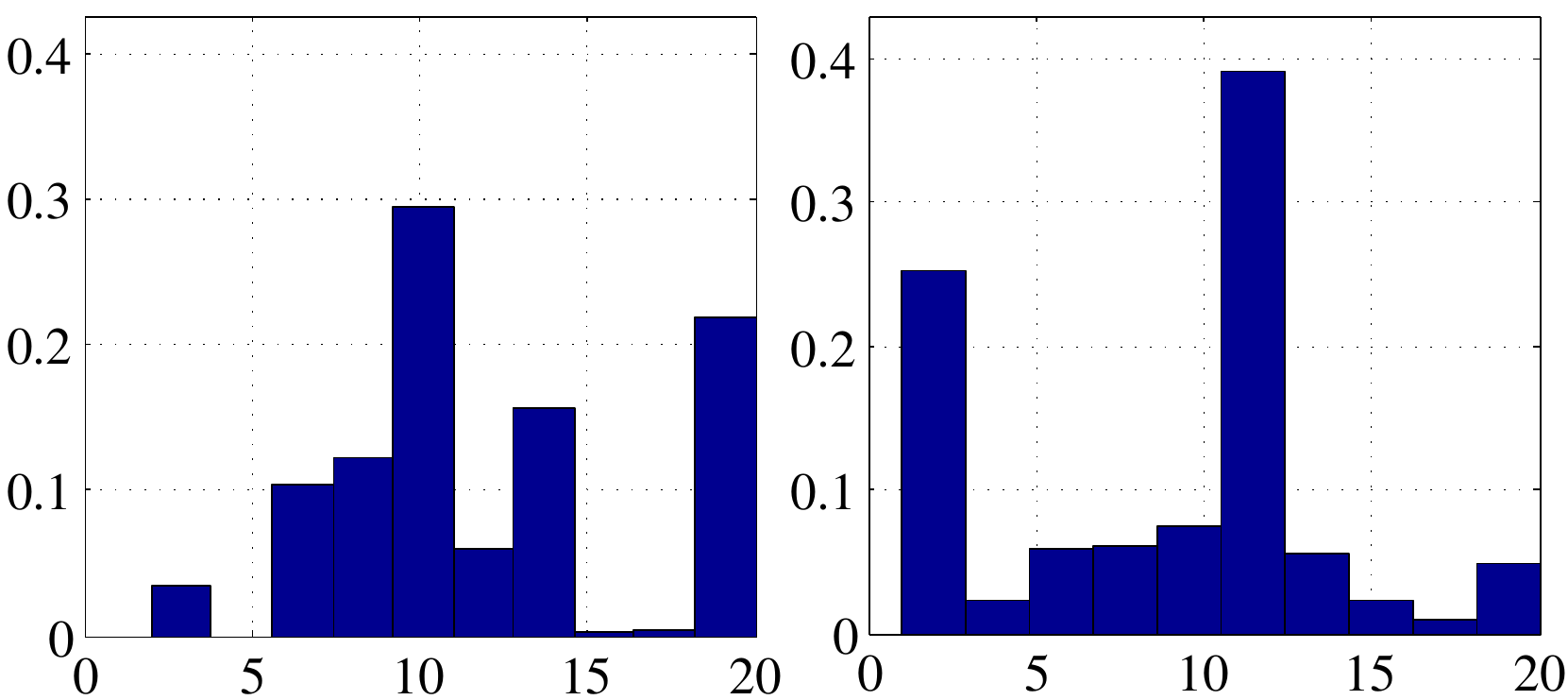}

\caption{Normalized histograms of the state variable values associated with data labeled as adherence (left) and non-adherence (right) to the walking test protocol.} 
\label{fig.HistogramIndicators}
\end{figure}

The multinomial naive Bayes is linear in the log-space of the input
variables, making it very easy to understand; we demonstrate this
by plotting a projection of the input variables and the decision boundary
in 2-D (Figure \ref{fig.LDAprojection}). The naive Bayes classifier
requires very little training data to estimate parameters, scales
linearly with the data size, and despite its simplicity has shown
performance close to state of the art for demanding applications such
as topic modeling in natural language processing, spam detection in
electronic communications and others \citep{Hand2001}. One of the
main disadvantages of this classifier is that it assumes, usually
unrealistically, that the input variables are independent, however
this is not an issue in this application since the classifier is trained
on a single feature. The multinomial naive Bayes classifier assumes
that data in the different classification classes follow different
multinomial distributions (Figure \ref{fig.HistogramIndicators}). 

For different clinimetric tests, we need to train different classifiers
because when the test protocols change so does the association between
the $z$'s and the $u$'s. However, the overall framework we use remains
universal across the different tests and can be extrapolated to handle
quality control in a wide set of clinimetric testing scenarios.

\section{Evaluation}

In order to evaluate the performance of the proposed framework, data
from 300 clinimetric tests (100 walking, 100 balance and 100 voice
tests) performed by PD patients and healthy controls from the Smartphone-PD
study (see \citet{abiola2015smartphone}) was processed using the
steps described above. The accelerometer data from the walking and
balance tests is recorded at frequency rates varying between 50Hz
and 200Hz. It is interpolated to a uniform rate of 120Hz (using standard
cubic spline interpolation) and the orientation is removed using $L_{1}$-trend
filtering as described in Section \ref{subsec:Isolating-and-removing}.
We extract the log-amplitude of the dynamic acceleration component
for the walking tests and the amplitude of the dynamic component for
the balance tests which serve as input to the behavioural segmentation
step. The log-amplitude for the walking tests is also down-sampled
by factor $4$ resulting in a length of $\sim90,000$ one-dimensional
sequential preprocessed time series; balance tests are not down-sampled
giving a length of $\sim241,000$ one-dimensional time series. For
the voice tests, the extracted energy of each 10 ms frame consists
of a length of $\sim200,000$ one-dimensional energy time series.

First, the 2-component GMM-based approach described above is evaluated
for each of the three data sets where performance is reported in Table
\ref{tab:Performance of the framework}. The metrics are estimated
using the expressions \eqref{eq:Accuracy metrics} where we compare
the estimated binary indicators $\hat{u}_{1},\dots,\hat{u}_{T}$ and
``ground truth'' labels $u_{1},\dots,u_{T}$ denoting adherence/violation.
Since this approach is completely unsupervised, we use all the data
for training, but none of the labels are used for validation. While
the GMM is not flexible enough to model the full composite behavioural
complexity of the data captured during most clinimetric tests, we
observe high accuracies for the voice tests. This is because the protocol
for voice tests consist only of producing sustained vocal phonations
in very close proximity to the sensor. Therefore, it can be argued
that adherence to the protocol for this activity is distinguishable
based on an appropriate measure of magnitude of the sensor recordings
alone, largely ignoring the longer-scale temporal variations. In tests
where the protocol requires the user to perform behaviours with more
complex, composite dynamics, the limitations of the simple GMM become
more apparent. For example, this occurs during the walking tests where
protocol violations can be distinguished a lot more accurately if
the behavioural segmentation model also incorporates both the longer-term
sequential nature of the data and its spectral information.

Next, the nonparametric switching AR behavioural segmentation model
is fitted to each of the three data sets where we specify the maximum
order of the AR models associated with each state to $r=4$. For the
evaluation here parameter inference is performed using the truncated
block Gibbs sampler described in \citet{Fox2009}. For future, real-time
deployment we can use more scalable deterministic algorithms based
on extensions of \citet{hughes2015scalable} and \citet{Raykov2016MAP}. 

As described above, we now input the state indicators $z_{1},\dots,z_{T}$
to a multinomial, naive Bayes classifier where if an indicator value
has not been seen during training, we assume it is classified as a
violation of the protocol during testing. The ability of the method
to correctly classify adherence and violations to the protocols of
the three tests is measured using standard 10-fold cross validation.
The mean and standard deviation of the BA, TP and TN rates of the
classifier are shown in Table \ref{tab:Performance of the framework}.
Note that in contrast to Section \eqref{subsec:Segmentation-context-mapping}
the accuracy metrics are now comparing the binary labels $u$ and
estimated $\hat{u}$. Both the TP and TN values are consistently high
across all tests, but the fact that the TN values are close to 90\%
for all three tests suggests a low probability of incorrectly labeling
data that adheres to the test protocol as a violation of the protocol.
In practice, the confidence in this prediction of adherence/violation
of protocol can be assessed using the state assignment probabilities
in the naive Bayes classifier associated with each time series data
point\footnote{The state assignment probabilities for each class of the naive Bayes
are the terms inside the $\argmax$ operator in \eqref{eq:naive Bayes rule}
after normalization.}.

In order to ensure that the reported classification accuracy is due
to the meaningful segmentation produced by the nonparametric switching
AR, we also report the performance of a multinomial naive Bayes classifier
(Table \ref{tab:Performance of the framework}) trained on the shuffled
state indicators estimated via the nonparametric switching AR during
the segmentation stage. In this way, the classifier is trained on
identical data but with randomized association between the data and
the training labels (i.e. $\left\{ z_{1},\dots,z_{T}\right\} $ are
randomly permuted while keeping $\left\{ u_{1},\dots,u_{T}\right\} $
fixed). If the association between estimated state indicators and
training labels is an accurate representation we would expect a classifier
trained on the shuffled indicators to score balanced accuracy of around
50\%.

\begin{table}

\caption{\label{tab:Performance of the framework}}

Performance of the two proposed algorithms for quality control
of clinimetric data and comparison with randomized classifier. For
the naive Bayes and the randomized classifiers quality control predictions
over all time points are evaluated using 10-fold cross validation.
We report mean and standard deviation (in the brackets) of the balanced
accuracy (BA), true positive (TP) and true negative (TN) rates across
the different cross validation trials changing the subsets of data
used for training and testing. For the GMM-based approach we report
the BA, TP and TN rates using all the data for training and for testing
since this approach is completely unsupervised; standard deviation
is not meaningful for a single trial (hence standard deviations are
omitted).
\par
\begin{centering}
\smallskip{}
\par\end{centering}
\centering{}%
\begin{tabular}{|c|c|c|c|}
\hline 
 & Walking tests & Balance tests & Voice tests\tabularnewline
\hline 
\hline 
\multicolumn{4}{|c|}{Naive Bayes + nonparametric switching AR }\tabularnewline
\hline 
BA & 85\% (11\%) & 81\% (14\%) & 89\% (8\%)\tabularnewline
\hline 
TP & 85\% (18\%) & 81\% (16\%) & 88\% (9\%)\tabularnewline
\hline 
TN & 90\% (8\%) & 88\% (9\%) & 91\% (9\%)\tabularnewline
\hline 
\multicolumn{4}{|c|}{GMM + running median filtering }\tabularnewline
\hline 
BA & 62\% & 24\% & 99\%\tabularnewline
\hline 
TP & 80\% & 74\% & 86\%\tabularnewline
\hline 
TN & 89\% & 82\% & 96\%\tabularnewline
\hline 
\multicolumn{4}{|c|}{Randomized classifier }\tabularnewline
\hline 
BA & 50\% (1\%) & 50\% (0.2\%) & 53\% (24\%)\tabularnewline
\hline 
TP & 1\% (0.4\%) & 0.4\% (0.02\%) & 99\% (1\%)\tabularnewline
\hline 
TN & 100\% & 100\% & 6\% (23\%)\tabularnewline
\hline 
\end{tabular}
\end{table}

\section{Future work}

\subsection{Simultaneous multiple sensors}

\noindent The proposed framework is not limited to specific types
of clinimetric activity tests nor specific sensors. It was demonstrated
that it can be applied to voice and accelerometer data from three
different smartphone clinimetric tests, but the approach could be
easily generalized to different sensor-generated time series. For
example, it would be straightforward to preprocess multiple sensor
types obtained simultaneously during a test and then model these combined
sensors together to perform segmentation. This may increase the accuracy
of quality control.

\subsection{Practical, real-time deployment}

\noindent Mobile, sensor-based applications that can support research
into the detection and symptom monitoring of PD have already been
deployed on a large scale outside the lab. The iPhone-based mPower
app and associated study has enrolled over 10,000 study participants
since 2015. Incorporating the proposed system into PD research apps
such as mPower could provide automatic, real-time quality control
of the smartphone clinimetric tests. This could automatically remove
any data which violates the test protocol, eliminating the need to
store, transfer or analyze unwanted data. Alternatively, data assessment
in real-time on board the data collection device could notify participants
to re-do a given clinimetric test to ensure that sufficient amounts
of useful, high quality data can be reliably collected.

The most computationally demanding component of the proposed system
is the inference of the nonparametric switching AR model. However,
the switching AR model can be seen as an extension of the infinite
HMM. Therefore in order to perform real-time inference of the model
on smartphone or wearable devices, we can use the computational optimization
proposed in \citet{Raykov2016} and \citet{Leech2017} which enables
inference of an infinite HMM on a highly resource-constrained microcontroller
computing device.

\subsection{Contextual learning}

\noindent ``Passive monitoring'', where sensor data is captured in
an entirely ambulatory way under realistic conditions outside the
lab, provides a way to study PD symptoms objectively without interrupting
routine activities. The successful monitoring of such daily behavioural
details may provide unprecedented insight into the objective monitoring
of individuals living with PD. However, outside the lab we usually
have little information about the routine activities under measurement
unless other, simultaneous monitoring methods are used, such as video
recording. However, video monitoring of patients in their homes is
expensive and can impact the integrity of the data as it is highly
invasive; patient behaviour may be altered under the awareness of
video monitoring. In addition, without multiple cameras in each room,
it is not possible to follow patients at different locations which
means that videoing every daily activity a patient performs is impractical.
The system proposed here can be trained to recognize specific patient
activities and help researchers identify segments of the passive monitoring
data which are most relevant for subsequent analysis. For example,
consider patients being passively monitored using smartphones, where
researchers wish to assess the effect of some medication on symptoms
such as slowness of movement \citep{Pavel2007} or postural sway \citep{arora2014high}.
With very few labelled instances of the relevant behaviours, the system
proposed here can learn to identify gait and balance behaviours from
the continuous, passive sensor data which will assist the researchers
into objectively testing their hypothesis.

\section{Conclusion}

This report describes a unified algorithm framework for quality control
segmentation of clinimetric tests utilizing sensing technologies to
remotely monitor patient health outside the lab. Monitoring patients
in their natural environment allows for a more realistic and accurate
assessment of an individual\textquoteright s health, improves accuracy
of outcomes in clinical trials in response to therapy and reduces
hospital stay \citep{martin2005challenge}. Nonetheless, the unknown
nature of the conditions under which data is collected in this way
raises suspicions about the quality of the data and its interpretability.
Thus, creating a systematic, automated algorithmic approach which
analyses the quality of the data could lead to smartphones and wearable
devices taking a central role as tools for scientific data collection
and monitoring.

This report uses a semi-supervised, nonparametric switching AR model
combined with a simple classifier to extract segments of smartphone
sensor data that adhere to the assumptions of the appropriate clinimetric
test protocol. The feasibility of this approach was demonstrated by
applying it to different smartphone clinimetric tests and sensor data
types, achieving segmentation accuracies of up to 90\% at the resolution
of the downsampled data.

By extracting segments of the sensor data that adhere to the assumptions
of a test protocol, it is possible to strip the data of confounding
factors in the environment that may threaten reproducibility and replicability.

\section*{Acknowledgment}

This work was partially supported by the Michael J. Fox Foundation
{[}Grant ID: 10824{]}, UCB Pharma, and in part by a research grant
from the NIH (P20 NS92529). The authors gratefully acknowledge Andong
Zhan for developing the smartphone application used in this study.
The authors express their sincere gratitude to every individual who
participated in this study to generate the data used here. 

\bibliographystyle{abbrvnat2}

\ifCLASSOPTIONcaptionsoff \newpage{}\fi
\end{document}